\documentclass[a4paper,11pt]{article}
\usepackage[margin=2.5cm]{geometry}
\usepackage{setspace,graphicx,enumitem}
\usepackage{amsmath,amssymb,amsfonts,bm,url,hyperref}
\newcommand{\N}{\ensuremath{\mathbb{N}}}
\newcommand{\R}{\ensuremath{\mathbb{R}}}

\begin{document}
\onehalfspacing
\title{Advanced anneal paths for improved quantum annealing}
\author{Elijah Pelofske, Georg Hahn, and Hristo Djidjev}
\date{Los Alamos National Laboratory}
\maketitle

\begin{abstract}
    Advances in quantum annealing technology make it possible to obtain high quality approximate solutions of important NP-hard problems. With the newer generations of the D-Wave annealer, more advanced features are available which allow the user to have greater control of the anneal process. In this contribution, we study how such features can help in improving the quality of the solutions returned by the annealer. Specifically, we focus on two of these features: reverse annealing and h-gain. Reverse annealing (RA) was designed to allow refining a known solution by backward annealing from a classical state representing the solution to a mid-anneal point where a transverse field is present, followed by an ordinary forward anneal, which is hoped to improve on the previous solution. The h-gain (HG) feature stands for time-dependent gain in Hamiltonian linear ($h$) biases and was originally developed to help study freezeout times and phase transitions in spin glasses. Here we apply HG to bias the quantum state in the beginning of the annealing process towards the known solution as in the RA case, but using a different apparatus. We also investigate a hybrid reverse annealing/h-gain schedule, which has a backward phase resembling an RA step and whose forward phase uses the HG idea. To optimize the parameters of the schedules, we employ a Bayesian optimization framework. We test all techniques on a variety of input problems including the weighted Maximum Cut problem and the weighted Maximum Clique problem. Our results show that each technique may dominate the others depending on the input instance, and that the HG technique is a viable alternative to RA for some problems.
\end{abstract}

\section{Introduction}
\label{sec:introduction}
Commercial quantum computers from D-Wave Systems Inc.~\cite{D-WaveSystems2000QuantumToday} make it possible to obtain approximate solutions of very high quality of many important NP-hard problems, such as the Maximum Clique problem, Vertex Cover, Graph Partitioning, and Graph Coloring. Specifically, such devices allow one to use a physical process called \textit{quantum annealing} (QA) to minimize quadratic unconstrained binary optimization (QUBO) or Ising functions in $n \in \N$ variables, defined by
\begin{align}
    Q(x_1,\ldots,x_n) = \sum_{i=1}^n h_i x_i + \sum_{i<j} J_{ij} x_i x_j.
    \label{eq:hamiltonian}
\end{align}
In eq.~\eqref{eq:hamiltonian}, the variables $x_i$ are unknown, whereas the linear weights $h_i \in \R$ and the quadratic couplers $J_{ij} \in \R$ for $i,j \in \{1,\ldots,n\}$ are specified by the user to define the problem under consideration. We call eq.~\eqref{eq:hamiltonian} a QUBO problem if $x_i \in \{0,1\}$ and an Ising problem if $x_i \in \{-1,+1\}$. Both the QUBO and Ising formulations are equivalent~\cite{Djidjev2016EfficientAnnealing}. As shown in~\cite{Lucas2014}, many important NP-hard problems can be formulated as a minimization problem of the form of eq.~\eqref{eq:hamiltonian}.

Since the first generation of D-Wave annealers (called D-Wave One) was introduced in 2009, more and more advanced features have been added to newer D-Wave generations, allowing the user greater control over the anneal process. Those features comprise spin reversal \cite{optimizing-spin-reversal}, customized anneal schedules or anneal offsets for individual qubits \cite{dwave-technology}. One of the two latest additions include reverse annealing schedules, and so-called time-dependent gain in linear biases, abbreviated as \textit{h-gain}.

In this work, we study how the latter two techniques, reverse annealing (RA) and h-gain biasing (HG), can be used to improve the quality of a solution returned by D-Wave. Whereas improving a known (suboptimal) solution is the motivation behind RA, both methods actually allow one to plant an approximate solution, obtained either classically or with a quantum technique, which is sought to be improved during the anneal.

In RA, the annealer performs a \textit{backward anneal} starting from a classical state representing the initial (planted) solution to a mid-point where a transverse field is present, followed by an ordinary forward anneal. If the initial solution is close to the global minimum, it is hoped that entering the quantum phase via the backward anneal will allow the annealer to transition to a better minimum, thereby improving upon the known solution.

The HG feature was originally designed to study freeze-out points \cite{JohnsonNature2011} and phase transitions in spin glasses \cite{Harris2018}, and allows one to weigh the linear term in eq.~\eqref{eq:hamiltonian} in a time-dependent way. In this contribution, we show that we can use the HG feature to plant an initial solution as in the RA, but using only forward annealing. Assuming the Ising function from eq.~\eqref{eq:hamiltonian} has no linear term, we add a new, suitable linear term that works as a bias towards the known initial solution. The HG feature allows us to put maximal weight on the linear term at the start of the anneal. Over the course of the anneal, we can decrease the HG strength to zero, thereby allowing the annealer to explore different solutions in the neighborhood of the planted one. The precise methodology is introduced in Section~\ref{sec:methods}. We also consider the application of HG to problems whose Ising formulations do have linear terms, and investigate a type of hybrid schedule that combines both RA and HG, specifically, that has a backward phase resembling an RA step and a forward phase based on the HG idea.

The implementation of the methodologies investigated in this contribution for encoding an initial solution depend on a variety of tuning parameters. In particular,  for RA, we need to choose the total anneal time and the schedule parameters. Likewise, HG depends on the total anneal time, the schedule parameters, and up to two scaling constants (depending on the structure of the Ising model in eq.~\eqref{eq:hamiltonian}), which we use to bias the solution towards the initial state. To tune those parameters, we employ a Bayesian optimization framework \cite{Mockus1974}, and we give details on how this optimization is being performed. Moreover, we present the best anneal schedules we obtained in this paper as a guidance on how to use RA and HG to encode initial solutions. 

The article is structured as follows. After a brief literature review in Section~\ref{sec:litreview}, Section~\ref{sec:methods} introduces the two techniques we investigate to encode an initial solution prior to the anneal process. Precisely, we describe RA in Section~\ref{sec:reverse_annealing}, and we give details on how to transform an input Ising of the form of eq.~\eqref{eq:hamiltonian} such that it encodes an initial state using the HG feature in Section~\ref{sec:hgain}. Experimental results are given in Section~\ref{sec:experiments} for the weighted Maximum Cut problem (Section~\ref{sec:experiments_maxcut}) and the weighted Maximum Clique problem (Section~\ref{sec:experiments_maxclique}). The article concludes with a discussion in Section~\ref{sec:discussion}. 

\section{Previous work}
\label{sec:litreview}
Whereas the HG feature remains relatively unexplored, RA has been studied in greater depth by several authors. The idea of RA was first introduced in the paper of \cite{Perdomo2011study} under the name of \textit{sombrero adiabatic quantum computation} and tested on 3-SAT instances. The authors observed that the performance of their algorithm was largely determined by the Hamming distance between the planted initial guess and the optimal solution.

A variety of techniques to perform local searches in the neighborhood of specified states via repeated calls of a quantum device is examined in \cite{Chancellor2017}. However, the author only assumes that a quantum annealer can be called with an initial state and does not explicitly consider reverse annealing.

In \cite{Ohkuwa2018}, the authors introduce a theoretical framework to show under which conditions RA can lead to improvements over QA for the fully connected $p$-spin model. However, they remark that their results may not apply to experimental setups where RA is performed diabatically and in a thermal environment.

In \cite{Yamashiro2019} the authors analyze, using direct numerical
integration of the time-dependent Schr{\"o}dinger equation, two types of RA, adiabatic RA, which is a forward annealing similar to the HG version studied here, and iterative RA as used in the D-Wave annealer. They show that, in theory, adiabatic RA provides a speed-up over QA for solving the mean-field-type $p$-spin model, but conclude that iterative RA as used by D-Wave does not provide this advantage in theory. As in \cite{Ohkuwa2018}, the authors remark that D-Wave is not a closed system, and thus theoretical results may not apply.

An empirical study on portfolio optimization is presented in \cite{Venturelli2019}. The authors observe a considerable speedup of RA over QA when seeded with heuristic solutions.

Another empirical study of RA in the context of $3$-spin models is presented in \cite{Passarelli2020}. The authors show that the open system dynamics, in connection with pausing, allow RA to converge to the ground state with a higher success probability than observed for purely closed system RA. The authors conclude that it is the open system dynamics that makes RA work in machines such as D-Wave.

\section{Methods}
\label{sec:methods}
This section describes the techniques we use to encode an initial solution, both via the D-Wave's RA feature (Section~\ref{sec:reverse_annealing}) and with the help of suitable linear terms in connection with the HG feature (Sections~\ref{sec:hgain} and \ref{sec:hgain_linear}). Section~\ref{sec:ra+hg_method} describes our hybrid technique of RA+HG. Section~\ref{sec:parameters} briefly describes the Bayesian optimization framework we use to optimize parameters and schedules.

\subsection{Anneal paths based on reverse annealing}
\label{sec:reverse_annealing}

\begin{figure*}
    \centering
    \includegraphics[width=0.95\textwidth]{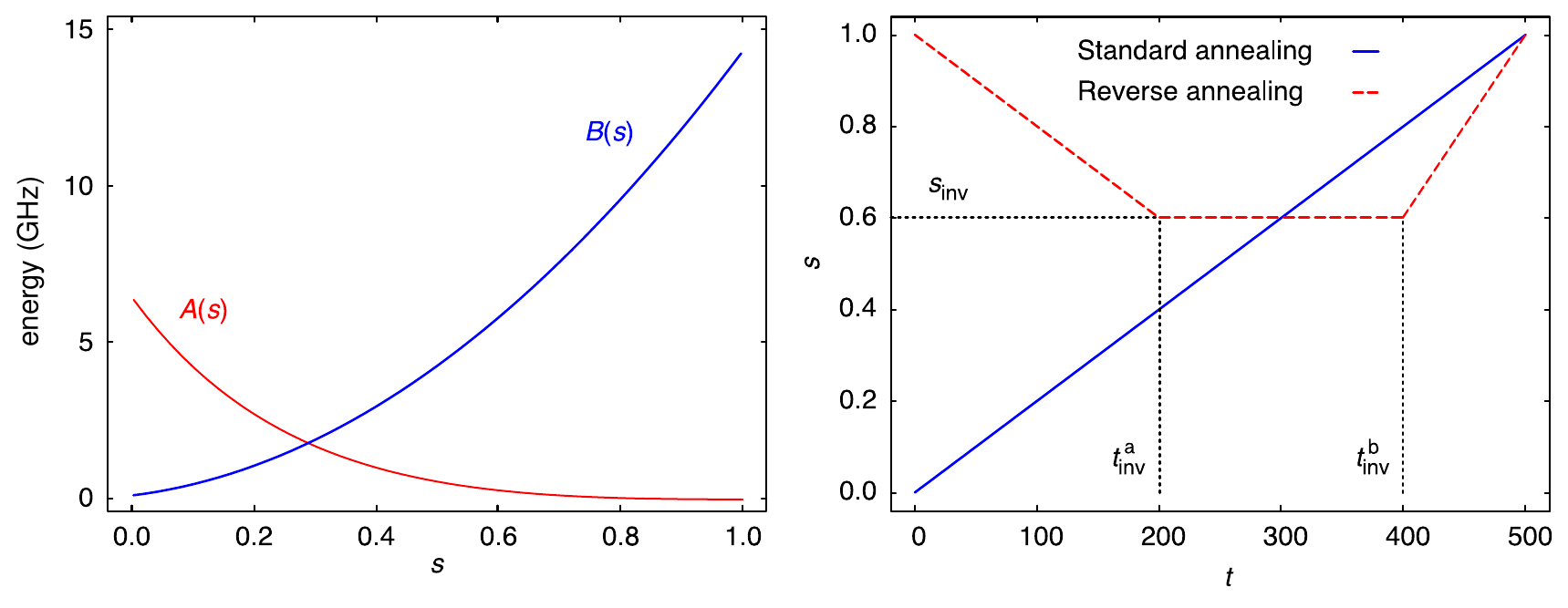}
    \caption{Left: Functions $A(s)$ and $B(s)$ controlling the anneal process, where $s \in [0,1]$ is the annealing fraction. Right: Progression of the anneal fraction $s$ for standard forward and reverse annealing with pause as a function of time $t \in [0,500]$ ms. Figure adapted from~\cite{Passarelli2020}.\label{fig:RA}}
\end{figure*}

In a standard forward anneal (FA), all qubits are prepared in an equal superposition of all   states, as determined by the transverse field portion of the system's Hamiltonian. During annealing, the amplitude of the transverse field is being decreased towards zero, while the Hamiltonian is slowly transformed into a Hamiltonian corresponding to the Ising problem being minimized. Specifically, the evolution of the D-Wave's quantum system is described by the following time-dependent Hamiltonian
\begin{equation}
    \label{eq:FA}
    H(s)=-\frac{A(s)}{2}\Big(\sum_{i=1}^n \hat{\sigma}^{(i)}_x\Big) +\frac{B(s)}{2}\Big(\sum_{i=1}^nh_i\hat{\sigma}_z^{(i)} + \sum_{i\leq j} J_{ij} \hat{\sigma}_z^{(i)} \hat{\sigma}_z^{(j)}\Big),
\end{equation}
where the first term having the prefactor $-A(s)/2$ is the transverse field and the term following the prefactor $B(s)/2$ is the Hamiltonian corresponding to the Ising model of eq.~\eqref{eq:hamiltonian} to be implemented on the annealer. The specific functions $A(s)$ and $B(s)$ used for the D-Wave 2000Q machine at Los Alamos are shown on Figure~\ref{fig:RA} (left). These functions are indexed by a parameter $s \in [0,1]$ called the \textit{anneal fraction}, which itself is a function $s(t)$ of the time. In the case of the FA, it is given as $s(t)=t/T$, where $T$ is the full anneal time. 

In contrast to FA, reverse annealing (RA) starts with a planted classical solution that is hoped to be much closer in quality to an optimal one than a random starting point. Then, a two-stage process is initiated, during which quantum fluctuations are first increased by reducing the anneal fraction from $s=1$ to a value $s_\text{inv}\in(0,1)$ at time $t_\text{inv}^a$ (the red curve in Figure~\ref{fig:RA}, right). After the turning point is reached, and after an optional pause until time $t_\text{inv}^b$, the anneal follows again the path of a standard forward anneal from $s_\text{inv}$ up to $s=1$ at  full anneal time $T$. Careful choices of the turning point and the initial state can lead to improvements in the solution compared to a forward anneal, see~\cite{Perdomo2011, King2018}.

\subsection{Anneal paths based on the HG feature}
\label{sec:hgain}
The feature of a \textit{time-dependent gain in Hamiltonian linear biases} allows the user to have more control of the anneal process by biasing linear terms of an Ising model with the help of a time-dependent function $g(t)$ as follows:
\begin{align}
    \nonumber
    H_{\mathrm{HG}}(s) = &- \frac{A(s)}{2} \Big( \sum_{i=1}^n \hat{\sigma}_x^{(i)} \Big)\\
    &+ \frac{B(s)}{2} \Big( \sum_{i=1}^n g(t) h_i \hat{\sigma}_z^{(i)} + \sum_{i>j} J_{ij} \hat{\sigma}_z^{(i)} \hat{\sigma}_z^{(j)} \Big),
    \label{eq:hgain}
\end{align}
see~\cite{Harris2018}. Compared to eq.~\eqref{eq:FA}, we see that the linear terms of the Ising model in eq.~\eqref{eq:hgain} are weighted with a function $g(t)$, specified by the user, which controls the time-dependent gain for the linear terms. In our implementation, we initialize the function with $g(0) \in [0,5]$ (5 being the largest value allowed for D-Wave 2000Q) and decrease it to $g(T)=0$ using up to $20$ points on the schedule. The specification of the HG feature is actually more general than the way we use it in this work. For instance, the function $g(t)$ may actually return values in $[-5,5]$, it does not need to be monotonic, there is a (machine-dependent) bound of $500$ for the slope between changes in the schedule, and a (machine-dependent) upper bound on the number of points determining the schedule~\cite{TechnicalDescriptionDwave}.

In this paper, we employ the HG feature to encode an initial solution at the start of the anneal process. Assume we are given an Ising problem of the type of eq.~\eqref{eq:hamiltonian} with no linear term, i.e., $h_i=0$ for all $i$. The idea lays in the observation that, for a fixed initial value $\bm{x^{(0)}} = (x_1^0,\ldots,x_n^0) \in \{-1,+1\}^n$, the minimum of the special Ising function containing only a linear term,
\begin{equation}
    \label{eq:H0}
    h(\bm{x}) = \sum_{i=1}^n (-x_i^0) x_i    
\end{equation}
for $\bm{x}=(x_1,\ldots,x_n)$, is precisely $-n$, and it occurs at $\bm{x}=\bm{x^{(0)}}$. Hence we can define $h_i=-x_i^0$ for $i=1,\dots,n$ and use a HG annealing schedule of the type of eq.~\eqref{eq:hgain}. By putting a large weight on the linear terms at the start of the anneal using the function $g(t)$, we bias the annealing solution towards our planted solution $\bm x^{(0)}$. Over the course of the anneal, the HG bias (the function $g(t)$ in eq.~\eqref{eq:hgain}) is decreased towards zero, thus allowing the anneal process to move away from the planted solution and to explore alternative ones in its neighborhood.

However, in order for this idea to work, the original Ising model may not have a linear term, so we can create our own linear term to encode the initial solution. For instance, Maximum Cut, Graph Partitioning, and Number Partitioning are such NP-hard problems without linear terms~\cite{Lucas2014}. Most Ising formulations of NP-hard problems, however, seem to have linear terms. Next we will show that even for such problems the HG approach can be applicable.

\subsection{Using HG for Ising problems containing linear terms}
\label{sec:hgain_linear}
For problems whose Ising formulations do have linear terms, we apply the following transformation to eliminate them. First, we homogenize the polynomial in eq.~\eqref{eq:hamiltonian} by converting the linear term into a quadratic one. This is achieved by introducing a new variable $z \in \{-1,+1\}$, which we call a \textit{slack variable}. The slack variable $z$ is multiplied with each linear term, thus transforming eq.~\eqref{eq:hamiltonian} into
\begin{align}
    Q'(\bm{x},z) = \sum_{i=1}^n h_i x_i z + \sum_{i<j} J_{ij} x_i x_j.
    \label{eq:H'}
\end{align}
Note that $Q$ can be recovered from $Q'$ by setting $z=1$. Then we can apply the method as discussed in Section~\ref{sec:hgain}. After the end of the annealing process, we ignore all solutions with $z=-1$. We can guide the anneal process to favor solutions with $z=1$ by using an appropriate HG bias (initial solution).

\subsection{Our hybrid method combining RA+HG}
\label{sec:ra+hg_method}
The ideas of RA and HG can actually be combined in a single D-Wave call. To be precise, given an initial solution $\bm x^{(0)}$ to be encoded, we first apply the methodology of Sections~\ref{sec:hgain} and \ref{sec:hgain_linear} to arrive at a new Ising model encoding $\bm x^{(0)}$. We then solve the new Ising model using an RA schedule, which specifies the anneal fraction $s$ as a function of time, combined with an HG schedule, which specifies the gain $g(t)$ as a function of time.

If the HG Hamiltonian computed in Sections~\ref{sec:hgain} and \ref{sec:hgain_linear} requires a slack variable $z$, we also need to supply an initial state for $z$ when running RA. In order to reinforce $z=1$, we indeed set $z=1$ in the RA initial state additionally to $\bm x^{(0)}$.

\subsection{Parameter setting}
\label{sec:parameters}
For the effective implementation of the RA and HG methods, we need to determine appropriate values for a set of parameters, some optional, others required.

For HG, optional parameters are the coefficients $h_i$ from eq.~\eqref{eq:hgain}, for which we have so far suggested only their sign in eq.~\eqref{eq:H0}. While choosing individual weights for each $h_i$ will result in highest accuracy, it is also the most difficult to accomplish and beyond the scope of this paper. Instead, we use a single coefficient $\alpha_1$ for $i=1,\dots,n$ and, in the case when we need to homogenize the input Ising model, another coefficient $\alpha_2$ for the new variable $z$. 

Combining the above, we encode an initial state using the Ising model
\begin{align}
    Q_\mathrm{final}(\bm{x},z)=\alpha_1\Big(\sum_{i=1}^n (-x_i^0) x_i\Big) - \alpha_2 z + Q'(\bm{x},z),
    \label{eq:H_final}
\end{align}
which is a function of $x_1,\ldots,x_n$ and $z$. The two scaling constants $\alpha_1$ and $\alpha_2$ allow us to control the strength with which the bias towards the initial solution and the condition that $z=1$ are enforced. If the Ising model under consideration in eq.~\eqref{eq:hamiltonian} does not have a linear term, no new variable $z$ is needed and thus $\alpha_2=0$ in eq.~\eqref{eq:H_final}.

Parameters that are required for both RA and HG are the schedule parameters. For RA, we need the values $t_\text{inv}^a$, $t_\text{inv}^b$, and $s_\text{inv}$, see Figure~\ref{fig:RA}, plus the total anneal time $T$ and for HG we need the function $g(t)$ given as a polygonal line subject to D-Wave's restrictions on magnitude, angles, and number of points. While there is some previous work that can be used as guidance for setting the schedule, in the case of HG there is no such previous work. Hence, we apply an optimization procedure for choosing the HG parameters and, in order to make a fair comparison between RA and HG, we use the same method for choosing the RA parameters.

We employ the following procedure for parameter setting. The tuning is done separately for the two problems that we study in more detail  in the experiments of Section~\ref{sec:experiments}, the Maximum Cut and the Maximum Clique problems, as follows:

\begin{enumerate}
    \item We first fix the anneal time $T$, and then the annealing schedule for RA. After having determined $T$, we fix the starting point $(t=0,s=1)$ and the end point $(t=T,s=1)$, see Figure~\ref{fig:RA} (right). As in Figure~\ref{fig:RA} (right), we decrease the anneal fraction $s$ to a point $(t_\text{inv}^a,s_\text{inv})$. We then allow for a pause, meaning we also allow a point $(t_\text{inv}^b,s_\text{inv})$ at the same $s_\text{inv}$. All in all, we need to determine for RA four parameters: $T, t^a_\text{inv}, t^b_\text{inv},$ and $s_\text{inv}$.
    \item Similarly, for HG, we first fix $T$ and then the schedule's end points, starting at $(0,5)$ and ending at $(T,0)$. We allow for one point in-between, $(h,t)$, where $h \in [0,5]$ and $t \in (0,1)$. Together, three parameters are required for HG, that is, $T, h, t$. Note that such a shape for an HG schedule is by no means optimal, but we want to keep the number of parameters smaller so we can have a more manageable search space. But before determining the schedule parameters, we first determine the best scaling factors $\alpha_1$ and $\alpha_2$ in eq.~\eqref{eq:H_final}. If the Ising model under consideration in eq.~\eqref{eq:hamiltonian} only has quadratic terms, homogenizing the polynomial is not necessary and we thus only need to find $\alpha_1$ in the Hamiltonian of eq.~\eqref{eq:H_final}. Otherwise, both $\alpha_1$ and $\alpha_2$ are determined.
    \item For the hybrid technique of RA+HG, after having determined the scaling constants $\alpha_1$ and $\alpha_2$ and the total anneal time $T$, we are left with five parameters determining the schedules: $t^a_\text{inv}, t^b_\text{inv},$ and $s_\text{inv}$ for RA, and $h, t$ for HG.
\end{enumerate}

For optimizing the parameters, we employ the Bayesian optimization tool of~\cite{bayesianopt}.
\textit{Bayesian optimization}~\cite{Mockus1974, Mockus1977, Mockus1989, Mockus2012} is a sequential optimization strategy to find the global optimum of a smooth function without the need for derivatives. Briefly, a uniform prior is put over the search space on which the function under investigation is defined. After querying a few first function evaluations, a posterior distribution is calculated which incorporates the obtained knowledge of the function evaluations (the data). Importantly, the posterior allows one to quantify the uncertainty in all unexplored areas, and it simplifies to a point mass at those locations where the function has been queried (and which are thus known exactly). Under suitable smoothness assumption on the function being optimized, the posterior allows to exclude areas which cannot contain the global optimum, and iteratively refining unexplored areas will result in a confidence region for the global optimum. An advantage of Bayesian optimization and the reason we chose it in this research is the fact that it also works with functions that are noisy, which is the case when the function is based on the energy values returned by a quantum annealer.

\section{Experimental analysis}
\label{sec:experiments}
This section reports on a variety of experiments conducted to assess the performance of both RA and HG, as well as the hybrid of RA+HG, for improving a planted solution. The experiments are divided into two subsections in which we investigate two important NP-hard problems, the weighted Maximum Cut problem (Section~\ref{sec:experiments_maxcut}) and the weighted Maximum Clique problem (Section~\ref{sec:experiments_maxclique}).

The structure of both subsections is identical: we first fix the scaling constants in eq.~\eqref{eq:H_final} for HG before we determine a suitable anneal duration for applying each of the three methods. Afterwards, we employ Bayesian optimization to determine the best anneal schedule, parameterized as described in Section~\ref{sec:parameters}. Once both the anneal duration and the anneal schedule are found for each of the RA, HG, and RA+HG methods, we evaluate all three techniques in terms of either the cut value (for the Maximum Cut problem) or the clique weight (for the Maximum Clique problem).

All experiments are carried out on Erd{\H o}s--R{\'e}nyi random graphs~\cite{ErdosRenyi1960} with probability/density parameter $p$, where $p \in \{0.1,0.2,\ldots,0.9\}$. Once the Ising model coefficients for the Maximum Cut or Maximum Clique problem are computed for each test graph, we embed it with {\tt minorminer} \cite{minorminer} using a chain strength value of $2$ and the default SAPI settings given by D-Wave.

Throughout the experiments, we employ the RA feature of D-Wave with the \textit{reinitialize state} option being switched on (the default choice), meaning that D-Wave reinitializes the planted state before each anneal is performed.

Moreover, we always run the {\tt bayes\_opt} tool of~\cite{bayesianopt} using the following parameter: the number of points for random exploration is set to {\tt init\_points=100}, the number of iterations for optimization is set to {\tt n\_iter=200}, and the noise level is set to {\tt alpha=0.01}. The parameter {\tt alpha} indicates to the optimizer how noisy the optimization landscape is. Since D-Wave samples are quite noisy, we observed that setting {\tt alpha} to a higher value, such as $0.01$, is favorable. However, we observe that large values of {\tt alpha} seem to cause an error in the optimizer, while smaller values lead to insufficient exploration of the optimization landscape.

\subsection{Weighted Maximum Cut problem}
\label{sec:experiments_maxcut}
This section focuses on the weighted Maximum Cut problem, defined as follows. Given an undirected graph $G=(V,E)$ with edge weights $w(e)$ for each edge $e=(u,v)$ connecting two vertices $u,v \in V$, we define a \textit{cut} to be any partition of $V$ into the disjoint union $C_1 \cup C_2$, where $C_1 \subseteq V$ and $C_2=V \setminus C_1$. The set of cut edges, called \textit{cutset}, is defined as $\mathcal{E}=\{ e=(u,v) \in E: u \in C_1, v \in C_2\}$ and its weight is $\sum_{e \in \mathcal{E}} w(e)$. The \textit{weighted Maximum Cut problem} asks to find a cutset of maximum weight. The Ising formulation of the weighted Maximum Cut problem is obtained by modifying the (unweighted) formulation in~\cite{Hahn2017ReducingBQ}, resulting in
$$Q_{cut}(\bm{x})=\sum_{(i,j) \in E} w((i,j)) \cdot x_i x_j,$$
where $x_i, x_j \in \{-1,+1\}$. Since the Ising formulation of the Maximum Cut problem does not have linear terms, no slack variable $z$ is needed in the Ising formulation in eq.~\eqref{eq:H_final}. The scaling constant $\alpha_1$ for HG in eq.~\eqref{eq:H_final} is determined in Section~\ref{sec:experiments_maxcut_parameters}.

In order to have a baseline truth for comparing RA, HG, as well as RA+HG we proceed as follows. We generate random graphs with $65$ vertices, edge probability $p \in \{0.1,0.2,\ldots,0.9\}$, and uniformly drawn edge weights in $(-1,1)$. After fixing $10$ of those graphs for each density as well as their embeddings on D-Wave, we perform $1000$ anneals of duration $1$ ms. The best solution among those anneals is then taken as the baseline. When testing RA, HG and RA+HG, all values we report are averages over those $10$ graphs. Moreover, we generate another set of $10$ graphs for each density to use as a validation set.

\subsubsection{Setting  scaling factors  and anneal time}
\label{sec:experiments_maxcut_parameters}
We start by determining a suitable choice of the scaling factor $\alpha_1$ in eq.~\eqref{eq:H_final} for the Maximum Cut problem using the Bayesian optimization.

As  fitness function for the optimization, we use the improvement in the maximum cut over the baseline. Each time the optimizer issues a call to the fitness function, we supply the average of $10$ problems optimized with either RA, HG, or RA+HG (depending on which one is optimized) using the parameter set probed by the Bayesian framework. The fitness value is then the average maximum cut improvement over the baseline. We make the fitness function dependent on three parameters, the scaling factor $\alpha_1$ as well as the parameters $(h,t)$ determining the HG schedule (see Section~\ref{sec:parameters}). For this experiment the anneal time is set to $1$ ms.

\begin{figure}
    \centering
    \includegraphics[width=0.5\textwidth]{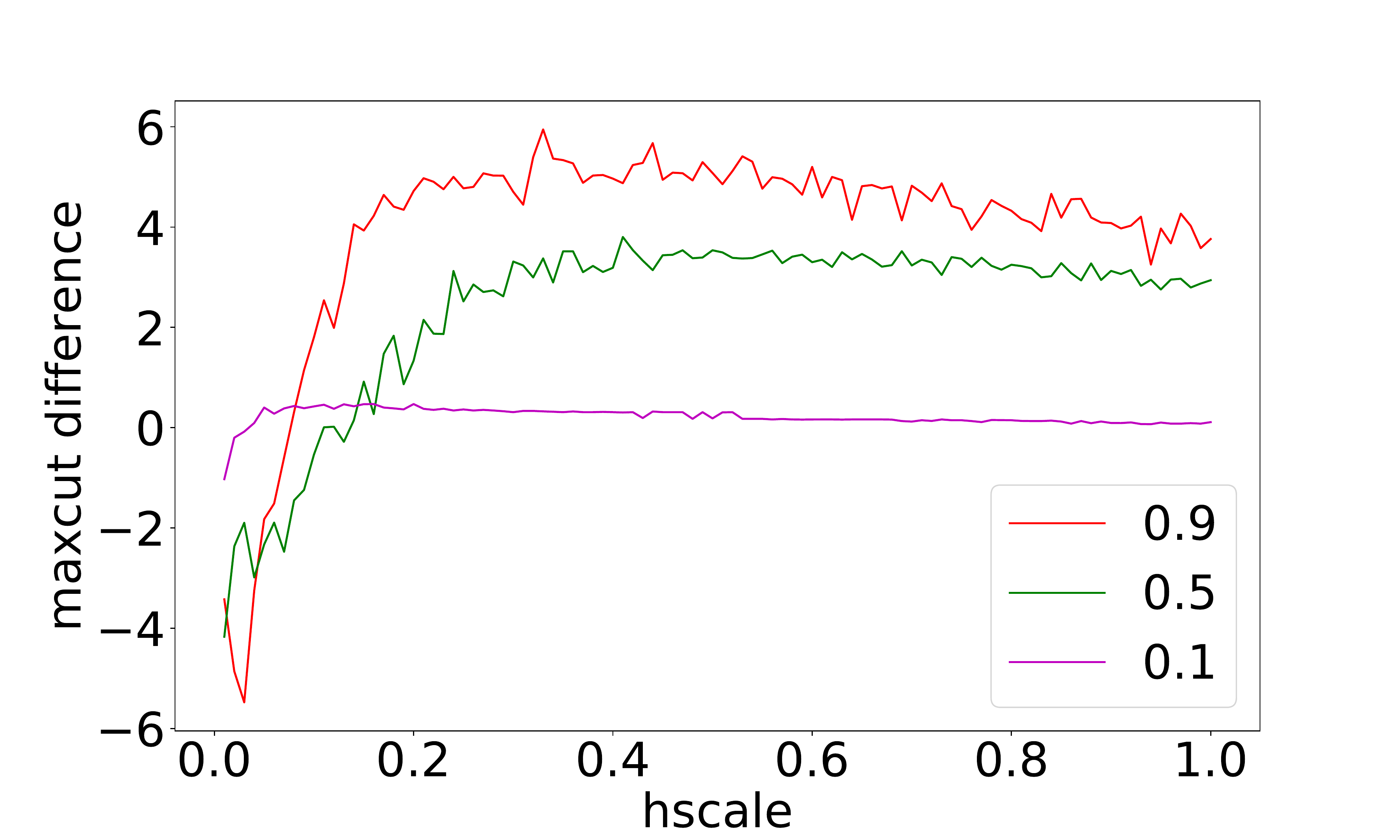}
    \caption{Maximum Cut problem. Difference in the maximum cut value to the baseline as a function of the HG scaling factor $\alpha_1$.}
    \label{fig:maxcut_scaling}
\end{figure}

After obtaining the fittest values, we fix the schedule $(h,t)$ and the anneal time of $1$ ms, and cross check the scaling factor $\alpha_1$ on a linear grid on $[0.01,1]$ in increments of $0.01$. Results for three different densities are shown in Figure~\ref{fig:maxcut_scaling}, which displays the difference in the maximum cut value to the baseline as a function of $\alpha_1$. We observe that the best choice of $\alpha_1$ is very dependent on the graph density, with e.g.\ the best choice for density $0.9$ occurring at $\alpha_1 \approx 0.3$. We will be selecting the scaling factor depending on the underlying graph density in the remainder of this section.

Next, we determine a suitable anneal duration for RA, HG, as well as the hybrid RA+HG. For this, we fix the HG schedule to the three points $[0, 5], [0.5T, 2.5], [T, 0]$ and the RA schedule to $[0, 1], [0.25T, 0.25], [0.75T, 0.25], [T, 1]$, where $T$ is the anneal duration (given in Table~\ref{tab:maxcut}). We note that these schedules are not optimal. Instead, they merely divide up the variable range in an equidistant fashion. We choose the anneal fraction to be around $0.25$ as suggested in~\cite{McGeoch2018}.

\begin{table*}
    \centering
    \begin{tabular}{l|c||ccccccccc}
        & anneal [ms] & 0.9 & 0.8 & 0.7 & 0.6 & 0.5 & 0.4 & 0.3 & 0.2 & 0.1\\
        \hline
        RA & 100 & 0.722 & -0.151 & 2.060 & 1.974 & 2.551 & 1.909 & 3.425 & 2.544 & 0.467\\
        RA & 2000 & 2.412 & -0.149 & 3.653 & 2.140 & 3.261 & 2.859 & 4.507 & 3.209 & 0.610\\
        HG & 1 & 5.173 & 2.084 & 3.72 & 2.632 & 3.127 & 2.799 & 2.876 & 2.082 & 0.338\\
        HG & 2000 & 3.963 & 1.266 & 2.216 & 1.421 & 2.047 & 1.617 & 1.705 & 1.525 & 0.185\\
        RA+HG & 100 & 3.853 & 2.832 & 4.867 & 3.022 & 3.478 & 3.170 & 4.32 & 2.726 & 0.526 \\ RA+HG & 2000 & 4.145 & 2.540 & 5.001 & 2.725 & 4.222 & 3.128 & 4.526 & 3.113 & 0.566
    \end{tabular}
    \caption{Evaluation of RA and HG, as well as hybrid RA+HG, for smallest and largest possible anneal times. Maximal cut difference on Erd{\H o}s--R{\'e}nyi graphs of density ranging from $0.1$ to $0.9$.\label{tab:maxcut}}
\end{table*}
Table~\ref{tab:maxcut} shows maximum cut results for the smallest and largest possible anneal times as a function of the graph density. We observe that an anneal duration of $2000$ ms works best for RA, while a $1$ ms anneal is best for HG. The hybrid technique of RA+HG does not seem to be as affected by the anneal duration, but since an anneal time of $2000$ ms yields slightly better results, we decide to employ RA+HG in connection with a $2000$ ms anneal in this section.

\subsubsection{Schedule computation via Bayesian optimization}
\label{sec:experiments_maxcut_bayes}
After having fixed the anneal duration for all three methods, we proceed by determining the parameters of the anneal schedule (see Section~\ref{sec:parameters}) via Bayesian optimization. For each density, we carry out a single run of the Bayesian optimizer.

\begin{figure}
    \centering
    \includegraphics[draft=false,width=0.5\textwidth]{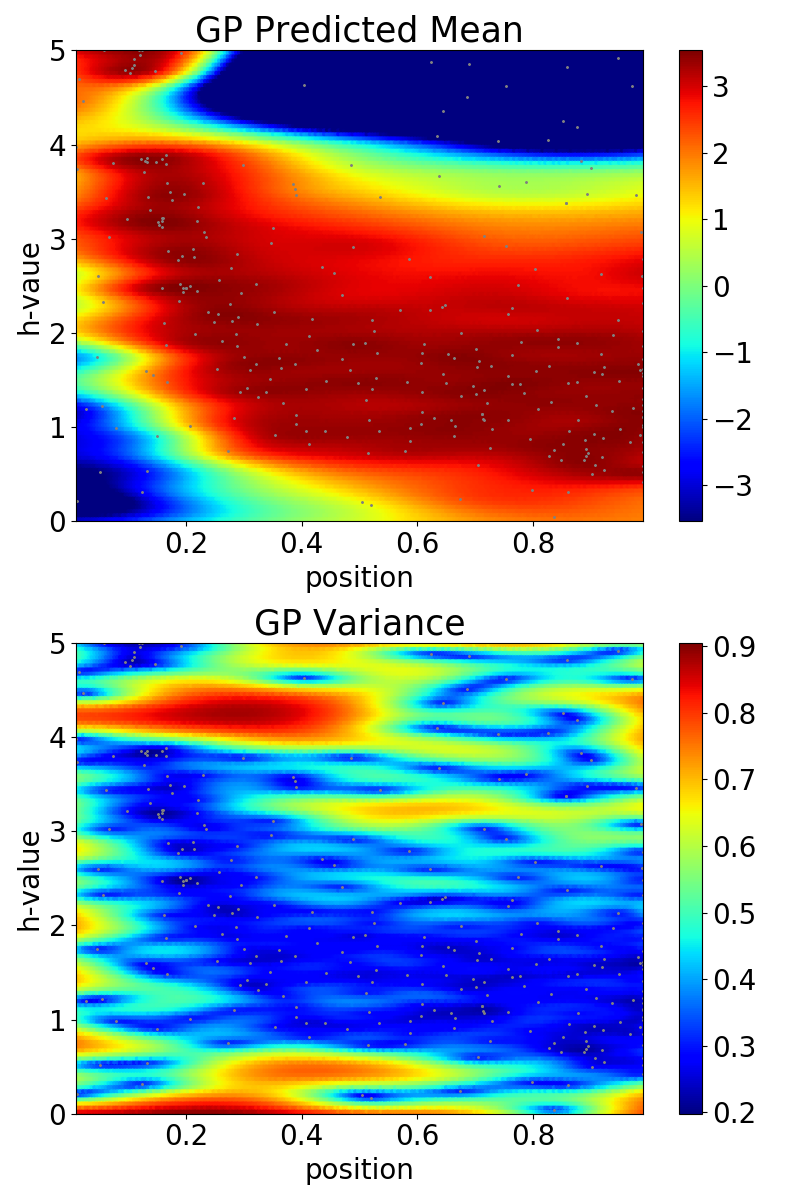}
    \caption{Bayesian optimization landscape for the HG schedule for Maximum Cut, visualized as a heatmap for graph density $0.5$ and anneal time $1$ ms. Top  shows maximum cut size improvement as a function of $g(t) \in [0,5]$ (see eq.~\eqref{eq:hgain}) on the y-axis, where $t$ is the position in the schedule (x-axis). Bottom  shows the variance of the Gaussian processes used by the Bayesian optimizer. Small dots indicate the points where the function was evaluated.}
    \label{fig:hgain_heat_map_maxcut}
\end{figure}
Since the schedule of HG has two parameters determining the midpoint in the anneal schedule (see Section~\ref{sec:parameters}), we can visualize its optimization as a heatmap in Figure~\ref{fig:hgain_heat_map_maxcut}. In particular, Figure~\ref{fig:hgain_heat_map_maxcut} shows the color-coded improvement in cut size over the baseline for each possible midpoint in the HG schedule. As described in Section~\ref{sec:parameters}, this point consists of a position in the anneal and a value of the HG function $g(t) \in [0,5]$, see eq.~\eqref{eq:hgain}.

The figure shows that the best choice of the HG value, defined as the one yielding the best improvement in maximum cut difference (red values), roughly decreases with the position in the anneal. We determine the maximum in this way for each density $p \in \{0.1,\ldots,0.9\}$. The schedules for RA (3 parameters) and RA+HG (five parameters) are fitted in a similar way, one schedule per density $p \in \{0.1,\ldots,0.9\}$.

\subsubsection{Comparing RA, HG, and RA+HG}
\label{sec:experiments_maxcut_comparison}
\begin{figure}
    \centering
    \includegraphics[width=0.5\textwidth]{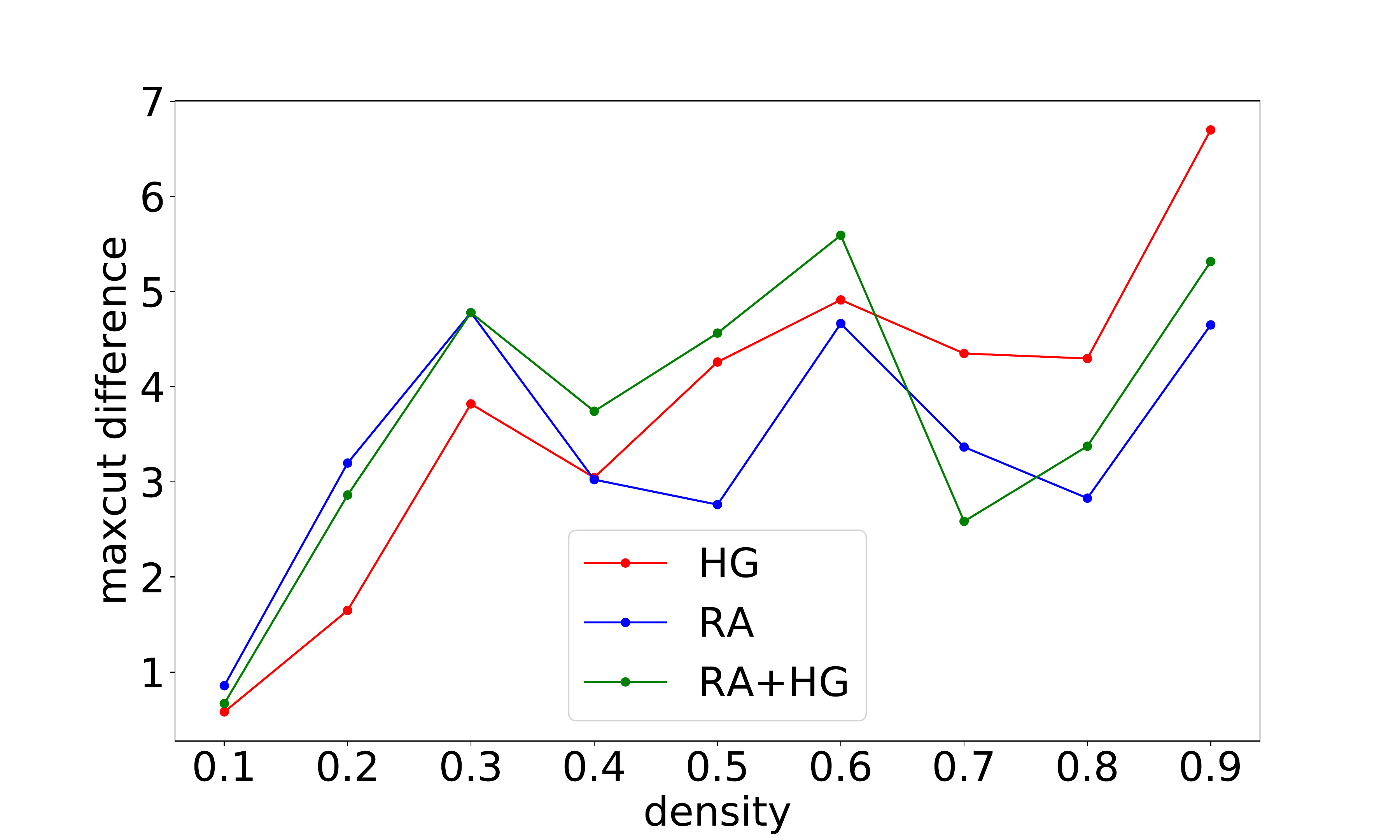}\hfill
    \caption{Comparison of RA, HG, and RA+HG with respect to the maximum cut improvement for the best schedules obtained via Bayesian optimization.}
    \label{fig:maxcut_densities_rerun}
\end{figure}

Having determined best schedule parameters for RA, HG, and RA+HG, we run again the experiment on 10 new graphs using these schedules. Figure~\ref{fig:maxcut_densities_rerun} shows results from this experiment. We observe that neither technique is uniformly better than the others. RA seems to be best for low densities, while HG and RA+HG perform best for high density graphs.

\subsubsection{Best schedules for RA, HG, and RA+HG}
\label{sec:experiments_maxcut_bestschedules}
It is interesting to look at the shape of some of the optimal schedules for RA and HG found by the Bayesian optimization. Additionally, we visualize one example of a schedule for RA+HG.

\begin{figure*}
    \centering
    \includegraphics[width=0.5\textwidth]{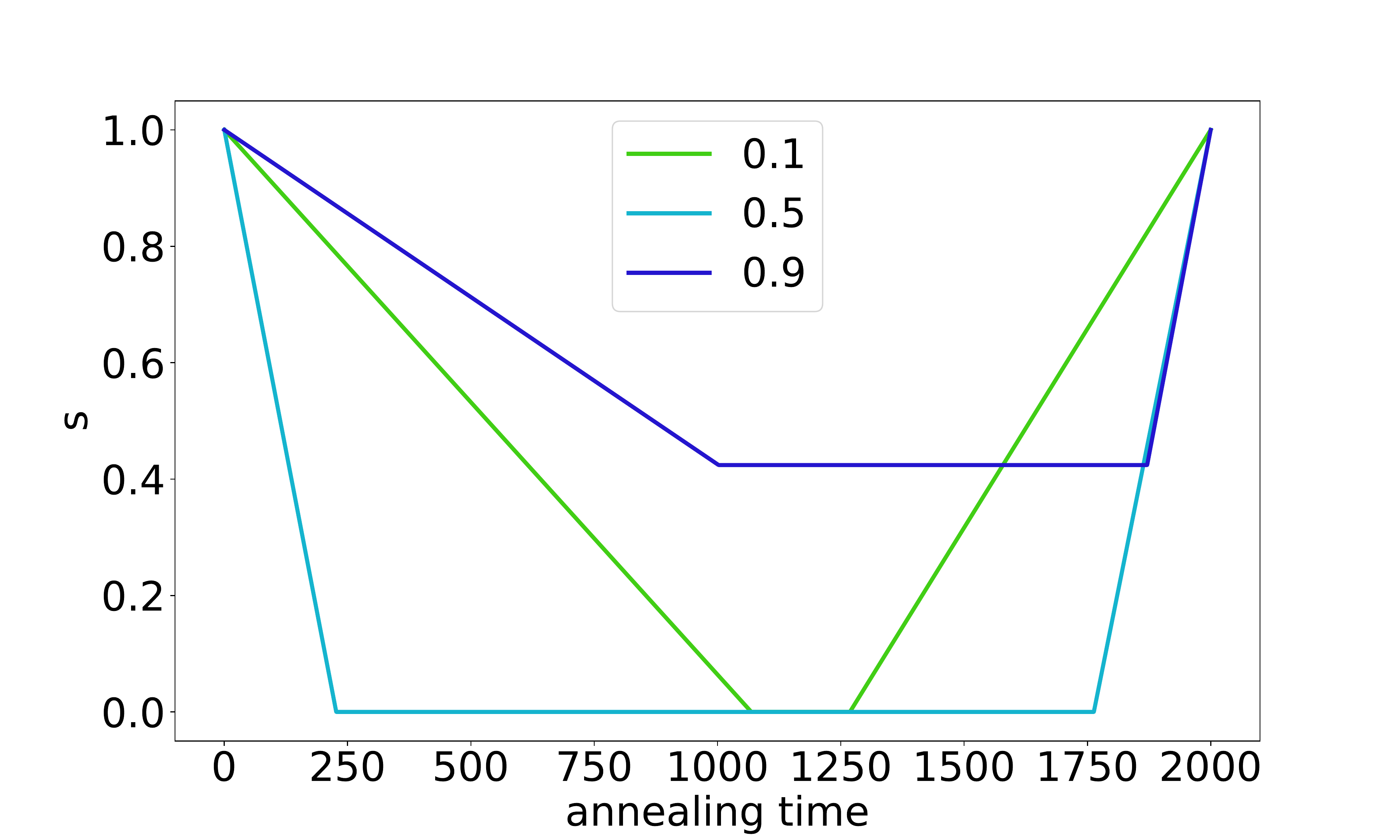}\hfill
    \includegraphics[width=0.5\textwidth]{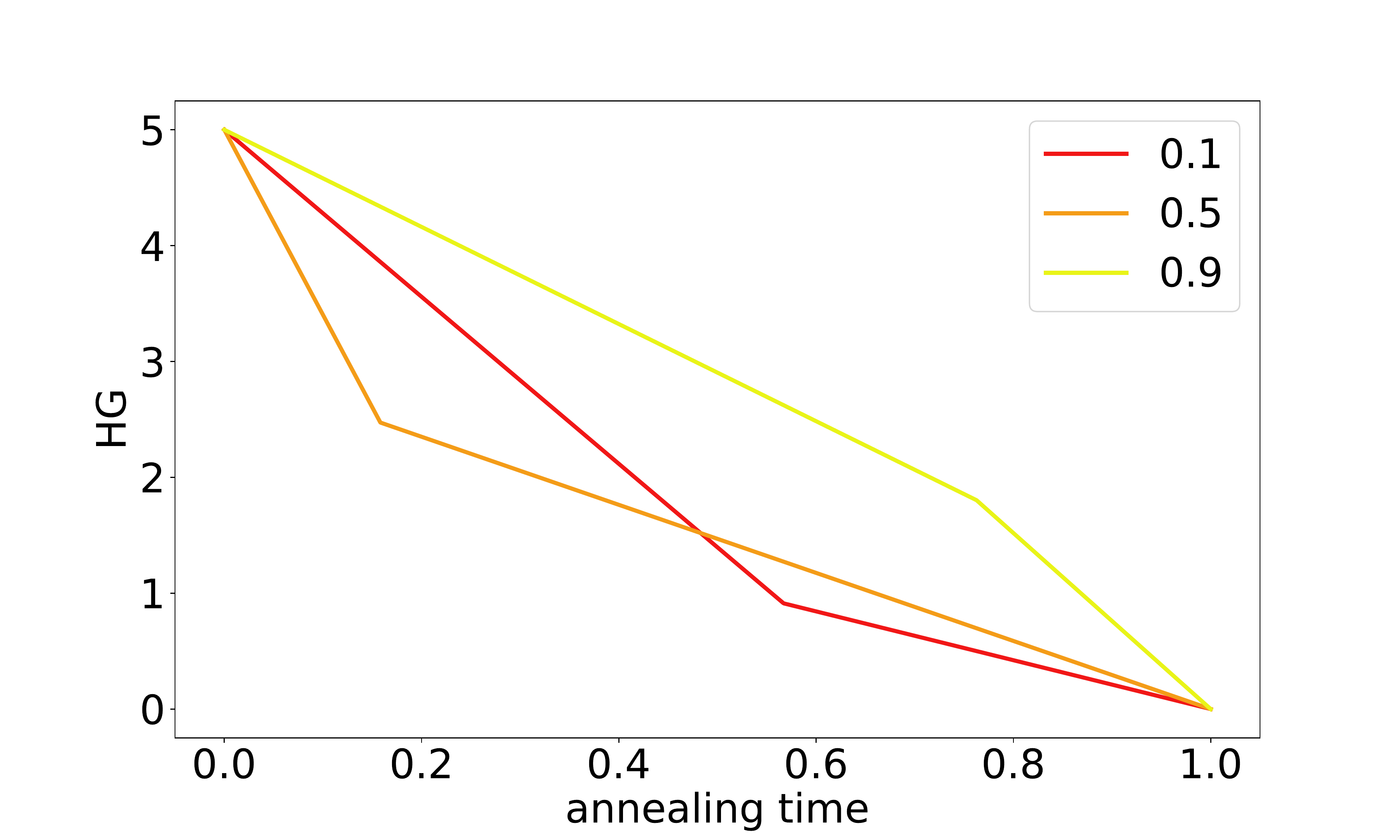}\hfill
    \caption{Maximum Cut problem. Best schedules for RA (left) and HG (right) for three different densities each, optimized for maximum cut difference. Each line is the best schedule for one density. }
    \label{fig:maxcut_schedules}
\end{figure*}

Figure~\ref{fig:maxcut_schedules} shows the best schedules for RA and HG color coded by density. For improved readability, we only display the schedules for $p \in \{0.1,0.5,0.9\}$.

We observe a pattern for the RA schedules in Figure~\ref{fig:maxcut_schedules} (left). In particular, when optimizing for maximum cut difference, RA schedules for low densities decrease down to an anneal fraction of zero, followed by a pause until roughly the midpoint of the anneal. In contrast, RA schedules for high densities only decrease to roughly an anneal fraction of $0.5$ at the midpoint of the anneal, followed by a pause until almost the full anneal time.

Similarly, a pattern can be observed for the HG schedules in Figure~\ref{fig:maxcut_schedules} (right). The HG schedules for low densities seem to have a steeper slope at the start of the anneal, and flatten off afterwards. In contrast, schedules for high densities seem to be closer to a straight line between the start point $(0,5)$ and the end point $(1,0)$.

\begin{figure}
    \centering
    \includegraphics[width=0.5\textwidth]{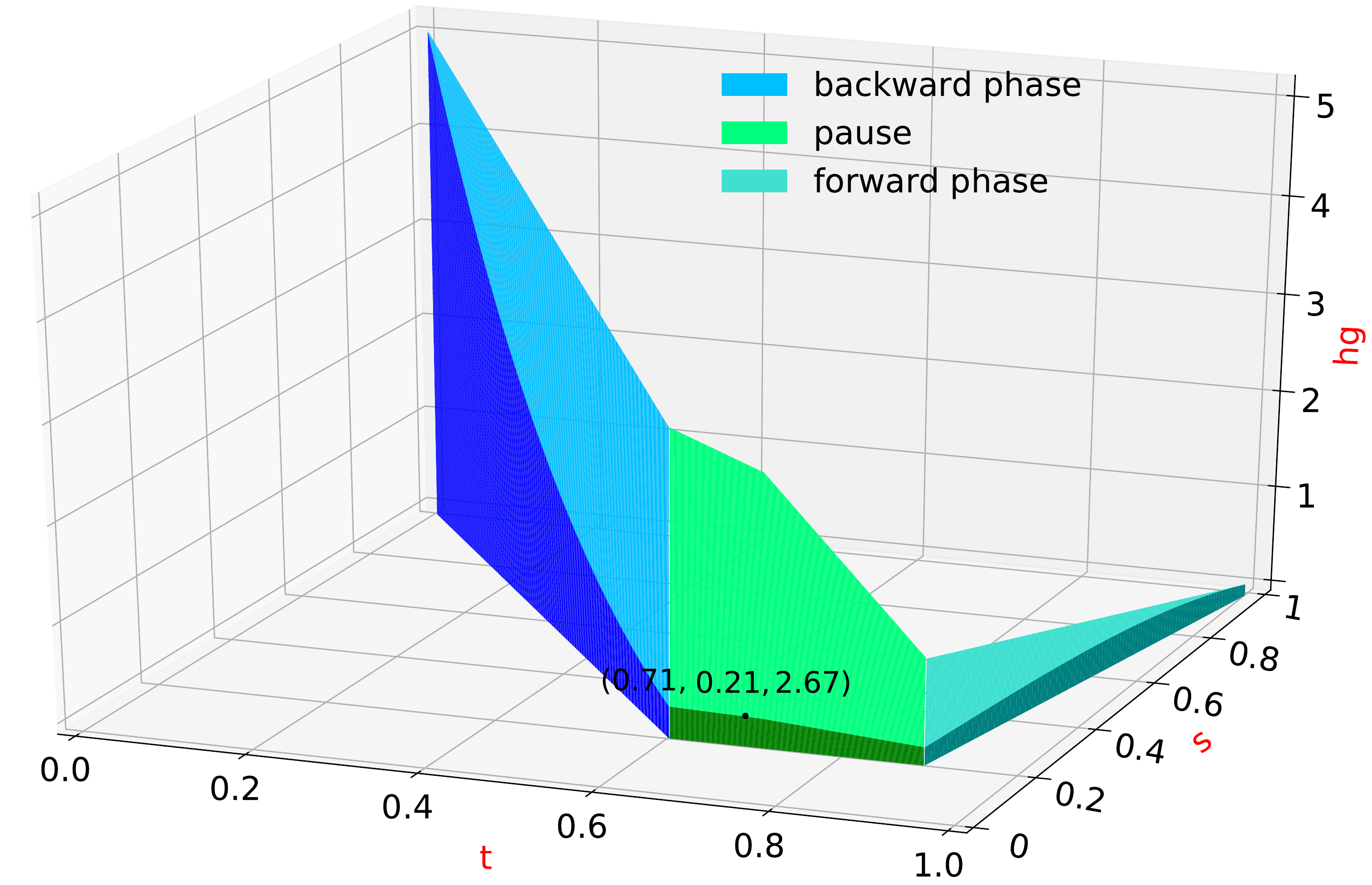}
    \caption{Illustration of an RA+HG schedule. Best schedule found for the Maximum Cut problem for $p=0.3$.}
    \label{fig:ra+hg_3d}
\end{figure}
An RA+HG anneal can be executed by sending to the D-Wave solver one RA schedule and one (independent) HG schedule. But while the RA aspect is easy to comprehend, the HG one is more difficult to grasp by just looking at the two component schedules because of the way the RA portion affects HG. Specifically, if $s=RA(t)$ and $g=HG(t)$ are the functions determined by the RA and HG schedules, respectively, then the real gain applied at time $t$ to the linear biases in eq.~\eqref{eq:hgain} is $(B(s)/2)HG(t)=B(RA(t))HG(t)/2$, where $B(s)$ is the function from eq.~\eqref{eq:FA}.
Figure~\ref{fig:ra+hg_3d} is given to help visualize the effect of an RA+HG schedule.
The RA component of the schedule, which has a pause for $t\in [0.6,0.89]$ at a value for $s$ equal to $0.21$, where $t$ is the time normalized in $[0,1]$, can be seen as a projection in the $t$-$s$ plane. The blue, green, and teal colors indicate the backward anneal, pause, and forward anneal phases. The HG schedule, which has middle point at $(t,hg)=(0.71,2.67)$, can be seen as the lighter-color projection in the $t$-$hg$ plane. Finally, the real gain applied to the linear biases at each time is represented by the darker-colored portion of the plot. The annotated point shows the value of the gain (2.67) at the middle point of the HG schedule (at 0.71). To simplify the plot, function $B$ has been normalized to $[0,1]$. We can see that the real gain applied during the pause and forward phases of the RA schedule stays mostly unchanged.

\subsection{Weighted Maximum Clique problem}
\label{sec:experiments_maxclique}
We carry out a similar analysis for the weighted Maximum Clique problem, defined as follows. For any graph $G=(V,E)$, a clique $C$ is a fully connected subset of vertices, i.e.\ $C \subseteq V$ such that $C \times C \subseteq E$. A maximum clique is a clique in $G$ of maximum size.

For the (vertex-)weighted version of the problem, we define a weight $w(v)$ for each vertex $v \in V$. The weight of a clique is accordingly defined as $w(C) = \sum_{v \in C} w(v)$. The weighted maximum clique problem asks for the clique $C \subseteq V$ having the largest weight $w(C)$. The QUBO formulation of the weighted Maximum Clique problem is obtained by modifying the (unweighted) formulation in~\cite{Chapuis2017}, resulting in
$$- \sum_{i=1}^n w(i) \cdot x_i + 2 \sum_{(i,j) \in E} \max \{ w(i),w(j) \} \cdot x_i x_j,$$
where $x_i,x_j \in \{0,1\}$. We can convert the above QUBO formulation into an Ising one using the equivalence given in \cite{Chapuis2019}. In contrast to the Maximum Cut problem investigated in Section~\ref{sec:experiments_maxcut}, the Maximum Clique formulation as an Ising model of the form of eq.~\eqref{eq:hamiltonian} does contain linear terms. We thus introduce a slack variable $z$ to homogenize the linear terms as in eq.~\eqref{eq:H'}, and add a new linear term encoding the initial solution as done in eq.~\eqref{eq:H_final}.

In the following experiments, we choose the vertex weights to be positive and randomly drawn in the range $(0.001,1)$.

\subsubsection{Setting scaling factors and anneal time}
\label{sec:experiments_maxclique_parameters}
We again focus first on the HG feature and repeat the tuning of Section~\ref{sec:experiments_maxcut_parameters}. In particular, to determine the two scaling factors $\alpha_1$ for $h(\bm x)$ and $\alpha_2$ for $z$, we run Bayesian optimization to fit both the schedule and the scaling factors simultaneously (see Section~\ref{sec:parameters}). For this, we fix the anneal time at $1$ ms.

Each time the Bayesian optimizer requests a new point, we return the average maximum clique improvement over the baseline (using $1000$ anneals) for $10$ graphs for each density. If no solutions are found, i.e.\ $z=-1$ for all $1000$ anneals, we return a large negative constant (we use $-1000$) to the optimizer.

After having obtained the result from the Bayesian optimization run, we fix the best schedule found. After initializing the Bayesian optimization algorithm with the parameters of the previous best solution (the previously found scaling constants $\alpha_1$ and $\alpha_2$ for the fixed schedule), we re-fit $\alpha_1$ and $\alpha_2$ with the help of the Bayesian optimization.

\begin{figure}
    \centering
    \includegraphics[draft=false,width=0.5\textwidth]{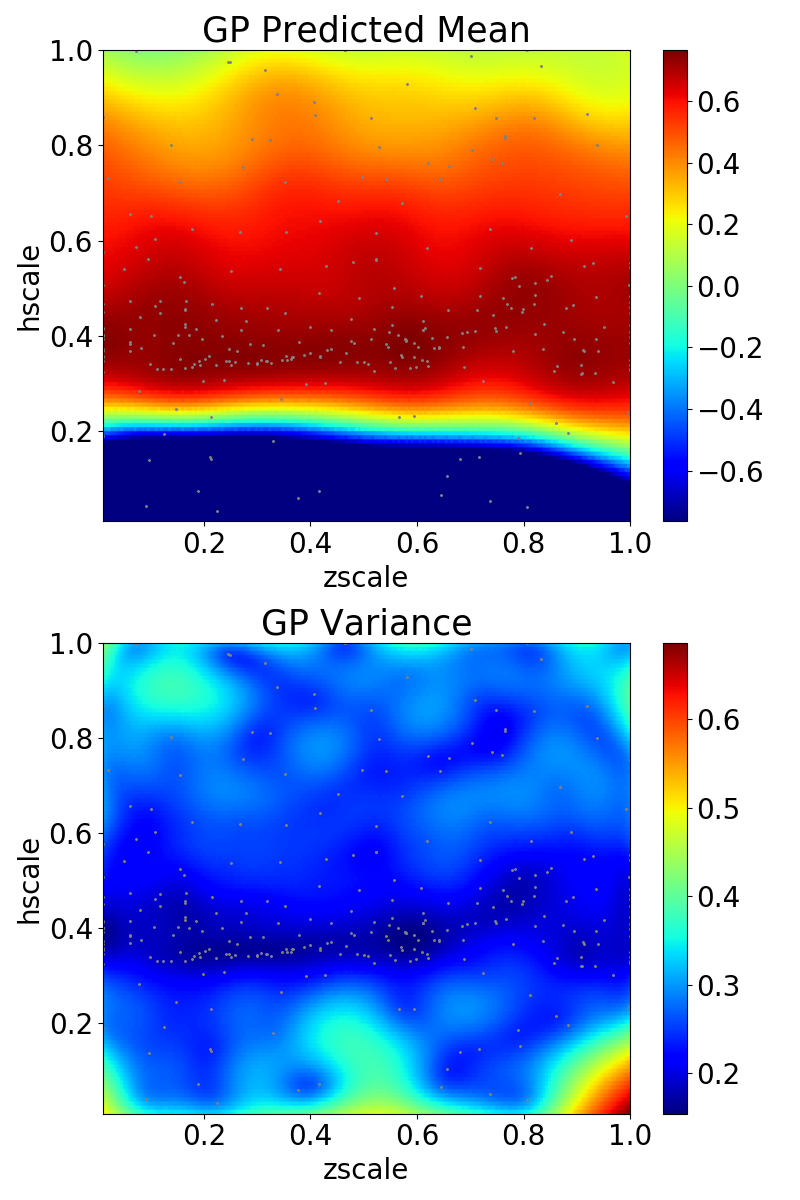}
    \caption{Landscape in $\alpha_1$ (h-scale on the y-axis) and $\alpha_2$ (z-scale on the x-axis) explored by Bayesian optimization. Graph density for $p=0.5$. Top plot shows mean of Bayesian posterior, bottom one shows variance (uncertainty).
    }
    \label{fig:scaling_factor_heat_maps}
\end{figure}
Figure~\ref{fig:scaling_factor_heat_maps} shows the result of the Bayesian optimization run with a fixed anneal duration of $1$ ms and a graph density of $p=0.5$, as well as our fixed optimized schedule. We see that the best values for the scaling constants are essentially in a band around $\alpha_1=0.4$ (h-scale), with various maxima for $\alpha_2$ (z-scale). The precise optimal scaling factors for $p=0.5$ returned by the Bayesian optimization are $\alpha_1=0.35$
(for h-scale) and $\alpha_2 = 0.25$ (for z-scale), which we fix for the remainder of this section.

We repeat this procedure for the other values of $p \in \{0.1,0.2,\ldots,0.9\}$ as well, and use individual scaling constants for each density in the remainder of this section as done for the Maximum Cut problem.

\begin{table*}
    \centering
    \begin{tabular}{l|c||ccccccccc}
        & anneal [ms] & 0.9 & 0.8 & 0.7 & 0.6 & 0.5 & 0.4 & 0.3 & 0.2 & 0.1\\
        \hline
        RA & 100 & 0.366 & 0.142 & 0.031 & 0.068 & -0.309 & -0.124 & -0.391 & -0.364 & -0.322\\
        RA & 2000 & 0.481 & 0.289 & -0.041 &  -0.060 & -0.322 & -0.171 & -0.474 & -0.408 & -0.309\\
        HG & 1 & 1.195 & 1.307 & 1.263 & 0 & 0 & 0 & 0 & 0 & 0\\
        HG & 2000 & 0.908 & 1.004 & 1.018 & 0 & 0 & 0 & 0 & 0 & 0\\
        RA+HG & 100 & 0.780 & -0.797 & -3.874 & 0.104 & 0 & 0.659 & 0.442 & 0 & 0 \\
        RA+HG & 2000 & 1.127 & 0.050 & -2.167 & 0.011 & 0 & 0.610 & 0.309 & 0.039 & 0
    \end{tabular}
    \caption{Evaluation of RA and HG, as well as hybrid RA+HG, for smallest and largest possible anneal times. Maximum clique difference on Erd{\H o}s--R{\'e}nyi graphs for densities ranging from $0.1$ to $0.9$.\label{tab:maxclique}}
\end{table*}
After having tuned HG, we now focus again on the three techniques (RA, HG and RA+HG). Similarly to Section~\ref{sec:experiments_maxcut_parameters}, we determine a suitable anneal duration for all three techniques by testing them for the shortest and longest anneal durations on Erd{\H o}s--R{\'e}nyi graphs of varying values of $p \in \{0,1,\ldots,0.9\}$.

Results are displayed in Table~\ref{tab:maxclique}, showing that RA works better for larger anneal durations (especially for denser graphs), and HG works consistently better for lower anneal durations. We will thus employ RA with an anneal time of $2000$ ms in the remainder of this section, and HG with an anneal time of $1$ ms. For RA+HG we fix the anneal duration at $2000$ ms.

\subsubsection{Schedule computation via Bayesian optimization}
\label{sec:experiments_maxclique_bayes}
The experiments of the previous sections allowed us to fix the anneal times, as well as the (density dependent) schedules and scaling constants for HG, RA, as well as RA+HG. Using the three calibrated techniques, we evaluate them on graphs of varying density with respect to the improvement of the maximum clique weight over the baseline.

\begin{figure*}
    \centering
    \includegraphics[width=0.5\textwidth]{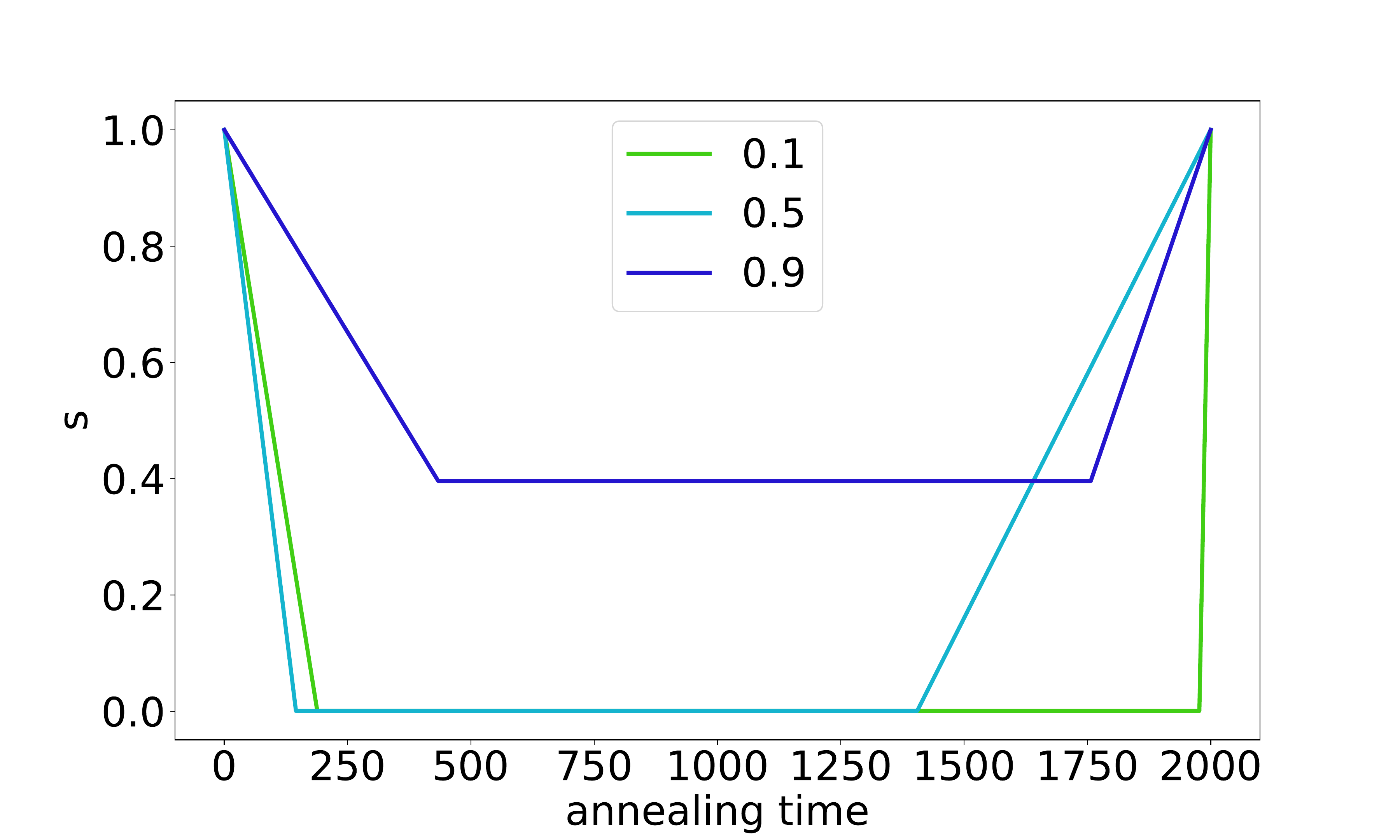}\hfill
    \includegraphics[width=0.5\textwidth]{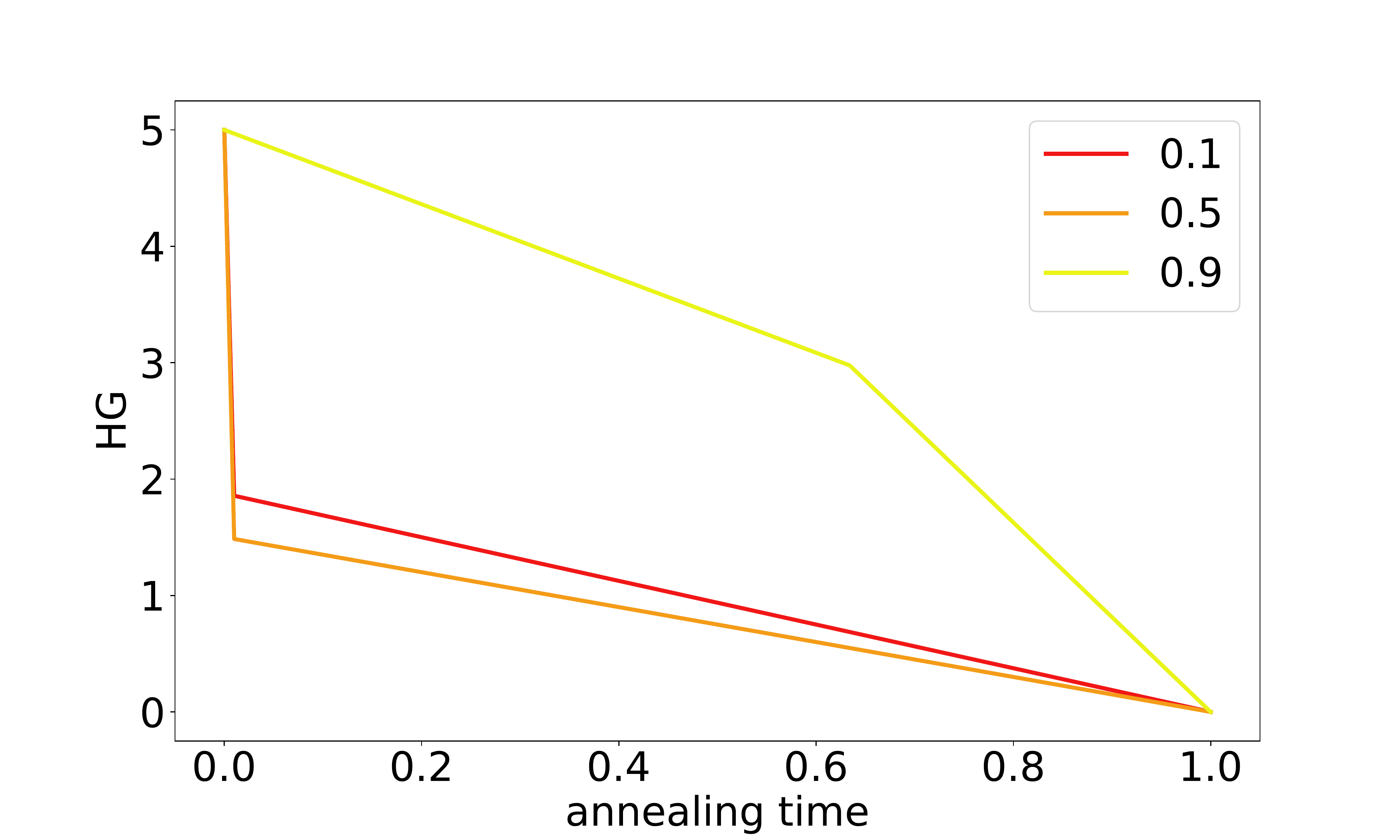}\hfill
    \caption{Maximum Clique problem. Best schedules for RA (left), and HG (right) optimized for maximum clique weight. Each line is the best schedule for one density.}
    \label{fig:maxclique_schedules}
\end{figure*}
Similarly to Figure~\ref{fig:maxcut_schedules} (left), we also report the best schedules found by the Bayesian optimization in the case of the Maximum Clique problem. All schedules are optimized to maximize the maximum clique weight over the baseline (a standard forward anneal).

Figure~\ref{fig:maxclique_schedules} shows the resulting schedules. For RA we see that, in contrast to the schedules for the Maximum Cut problem,  for low densities the optimal RA schedules decrease the anneal fraction at the start of the anneal to very small values, and perform a pause until almost the full anneal time. As the graph under consideration becomes denser, the anneal fraction is only decreased down to roughly $0.4$, and the pause occurs in the center having roughly a duration of half the anneal time. The schedules for HG (Figure~\ref{fig:maxclique_schedules}, right) resemble the ones observed for the Maximum Cut problem.

\begin{figure}
    \centering
    \includegraphics[draft=false,width=0.5\textwidth]{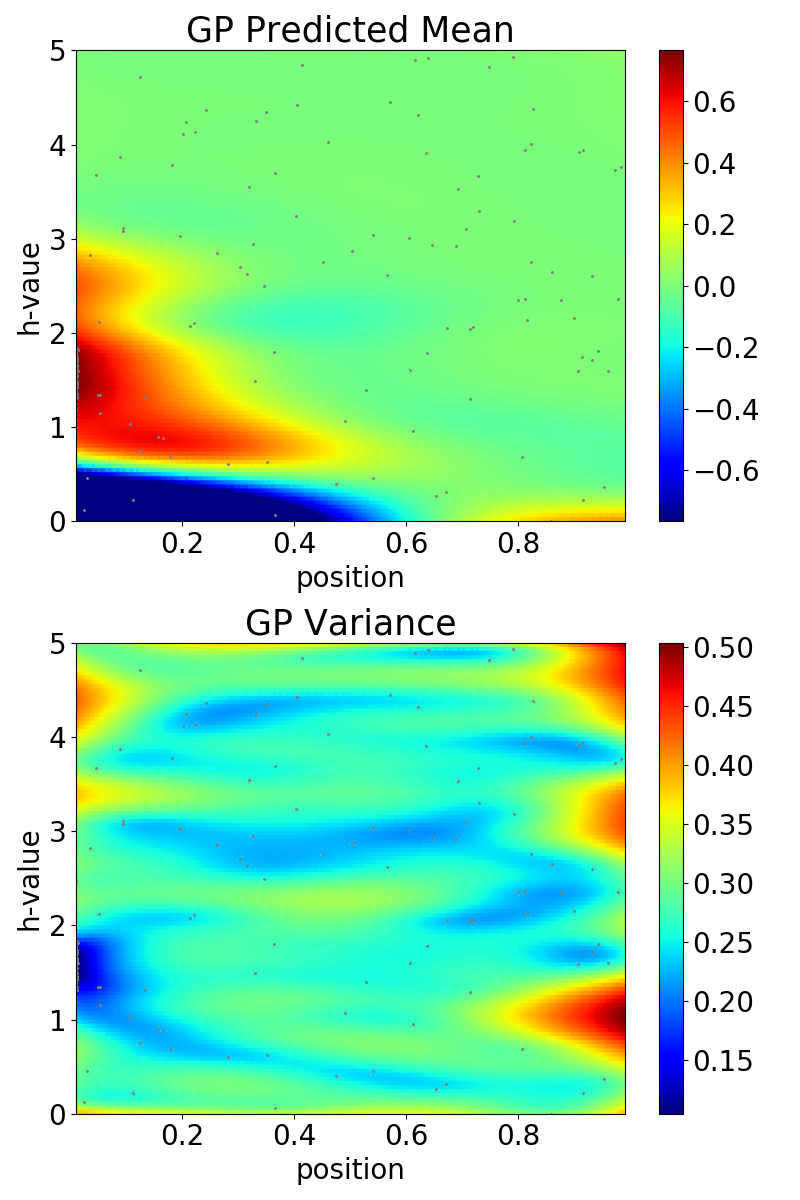}
    \caption{Maximum Clique problem. Bayesian optimization landscape for the HG schedule, visualized as a heatmap for $p=0.5$ and anneal time $1$ ms. Top shows maximum clique weight improvement as a function of $g(t) \in [0,5]$ (see eq.~\eqref{eq:hgain}) on the y-axis, where $t$ is the position in the schedule (x-axis). Bottom shows the variance of the Gaussian processes used by the Bayesian optimizer.}
    \label{fig:hgain_heat_maps}
\end{figure}
Similarly to Figure~\ref{fig:hgain_heat_map_maxcut}, we again visualize the optimization of the HG schedule as a heatmap in Figure~\ref{fig:hgain_heat_maps}. We see that the optimal point occurs at roughly position $0.2$ (anneal fraction) and has a HG value of around $1$.

\subsubsection{Comparing RA, HG, and RA+HG}
\label{sec:experiments_maxclique_comparison}
\begin{figure}
    \centering
    \includegraphics[width=0.5\textwidth]{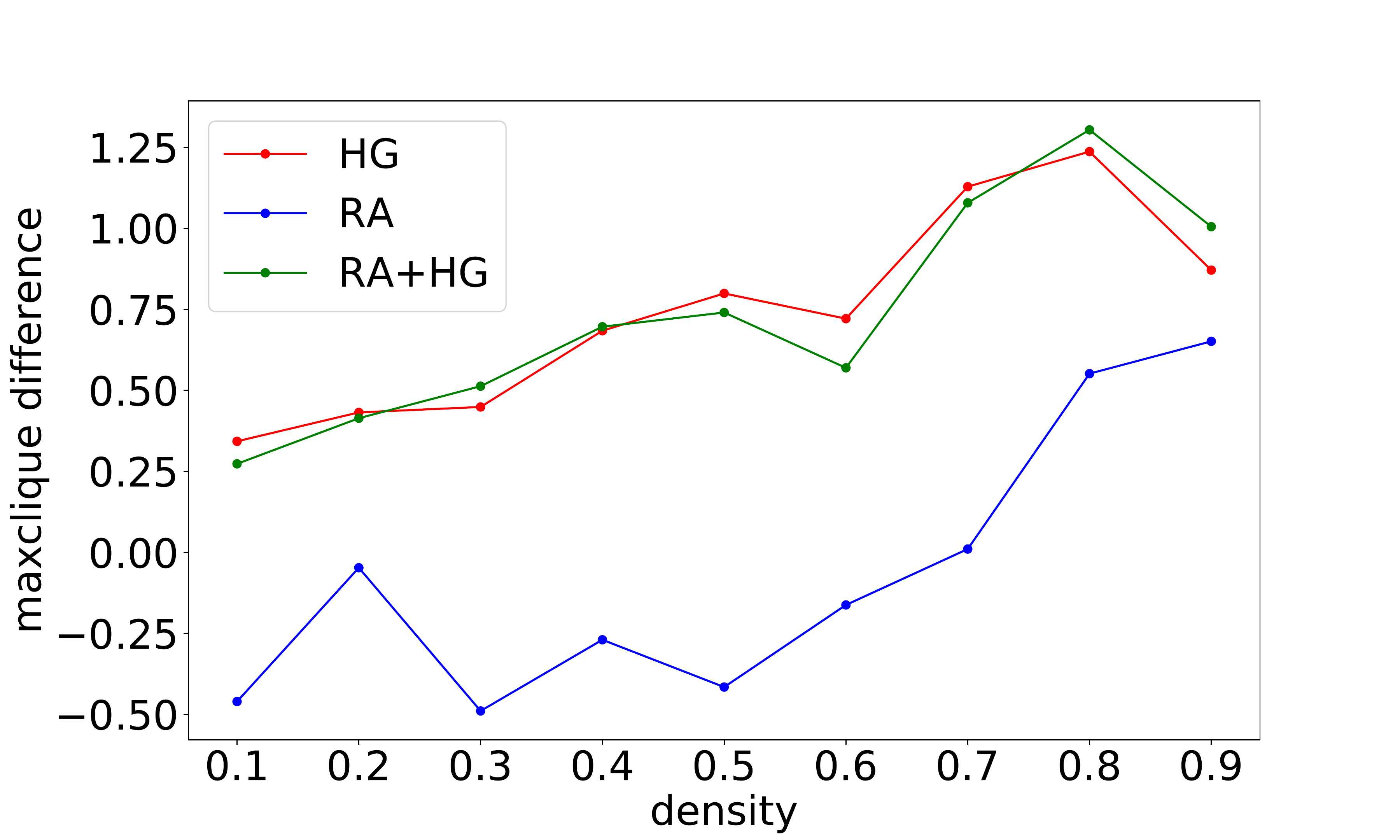}
    \caption{Comparison of RA, HG, and RA+HG with respect to the improvement in maximum clique weight over the baseline. Plot uses the $10$ new problems.}
    \label{fig:maxclique_comparisons}
\end{figure}
As in Section~\ref{sec:experiments_maxcut_comparison}, we evaluate RA, HG as well as RA+HG after tuning the scaling factors, anneal durations, and schedules. Results are shown in Figure~\ref{fig:maxclique_comparisons} for $10$ new problems not used in the training set. We observe that the behavior of all three techniques is consistent: On the new problems, RA performs worst with the exception of graph density corresponding to $p=0.9$. Both HG and RA+HG perform very similarly and consistently better than RA, although they draw equal with RA for  $p=0.9$.

This behavior is different from the equivalent experiment for Maximum Cut in Figure~\ref{fig:maxcut_densities_rerun}, where both HG and RA+HG were only marginally better than RA.

\section{Discussion}
\label{sec:discussion}
In this contribution we investigated two techniques suitable to encode an initial solution prior to the anneal on the D-Wave 2000Q quantum annealer. The two techniques are the native reverse annealing feature of the D-Wave device, as well as our own method based on the h-gain feature.

Since the two techniques rely on a variety of tuning parameters, we conduct extensive testing to determine suitable anneal times, parameters, and schedules. Using optimized sets of parameters we compare both methods on both the Maximum Cut problem (whose Ising formulation does not have linear terms, thus making our h-gain technique directly applicable), as well as the Maximum Clique problem (for which we have to transform the Ising model first). We summarize our findings as follows:
\begin{enumerate}
    \item The anneal durations for RA and HG seem to be very problem dependent. However, there is a consistent pattern in the anneal schedules for the two problems we considered: for graphs of lower density, RA schedules with an early and longer pause at a low anneal fraction are advantageous, whereas for higher densities a shorter pause at an anneal fraction of around $0.5$ seems better. For HG, the optimal schedules are close to the line connecting $(0,5)$ and $(1,0)$ independently of the density.
    \item The scaling constants can be found successfully via Bayesian optimization.
    \item In our experiments, the annealer almost always returned samples with value for the slack variable $z=1$ for the optimized HG schedules. Having $z=1$ is necessary for our technique to work, but we did not expect it to happen so often.
    The precise explanation for this observation is still unknown, but 
    one possible explanation is that the HG bias helps guiding the anneals towards solutions with $z=1$. 
    We also observe that $z=1$  occurs with much lower frequency for non-optimal HG schedules.
    \item We list several RA, HG, and RA+HG schedules for three optimization problems that other researchers can use in their codes if they solve problems from the classes studied here. But after examining  all these schedules, we were not able to discover any patterns to suggest simpler heuristic algorithms that can be used for finding good schedules for any new classes of problems, instead of running expensive optimizations for each new class. That does not necessarily mean that such heuristics do not exist, and finding such simpler rules can be an interesting problem for future research. 
    \item Overall, we conclude that our technique to plant initial solutions with the help of the HG feature, as well as RA+HG, seem to be a viable alternative to reverse annealing.
\end{enumerate}

This article leaves considerable scope for future work:
\begin{enumerate}
    \item In this work we only considered RA schedules with two points defining a pause, and HG schedules with one point. However, more complicated schedules for both RA and HG are possible, including other annealing times, and RA+HG schedules with more points.
    \item For the hybrid RA+HG technique, we encoded the same solution bitstring for both RA and HG. However, this is not necessary, and it remains to be investigated if two different initial solutions increase the quality of the solution after annealing.
    \item The technique based on HG we propose to encode an initial solution (Section~\ref{sec:hgain}) works for both Ising models without and with linear terms. We exemplarily show one candidate for each case, Maximum Cut having no linear terms, and Maximum Clique having linear terms. However, our HG technique can be applied to many more interesting problems such as graph partitioning, the traveling salesman problem, minimum vertex cover, or graph coloring. Additionally, many of those problems themselves exist in different variants, including unweighted, vertex- or edge-weighted formulations.
    \item We used the Bayesian optimization framework of \cite{bayesianopt} in a rather ad-hoc way. Tuning the parameters of the Bayesian optimization, in particular with the aim to make the optimization more robust against the noise in the D-Wave samples, could further improve the optimized parameters and schedules we report.
    \item For cases where $z$ does not always equal 1, one could observe if the proportion of anneals where $z=1$ is higher for RA+HG in comparison to HG only, assuming all other variables are held constant. This is conjectured to be the case because in RA+HG the value of $z$ is reinforced by the initial state of RA. 
    \item It would be interesting to consider iterative applications of the RA and HG methods where samples from one iteration are used to bias the annealing in the next one.
\end{enumerate}


\begin{thebibliography}{10}
\bibitem{Chancellor2017}
N~Chancellor.
\newblock Modernizing quantum annealing using local searches.
\newblock {\em New Journal of Physics}, 19(2):023024, 2017.

\bibitem{Chapuis2017}
G.~Chapuis, H.~Djidjev, G.~Hahn, and G.~Rizk.
\newblock {Finding Maximum Cliques on the D-Wave Quantum Annealer}.
\newblock In {\em Proceedings of the 2017 ACM International Conference on
  Computing Frontiers (CF'17)}, pages 1--8, 2017.

\bibitem{Chapuis2019}
Guillaume Chapuis, Hristo Djidjev, Georg Hahn, and Guillaume Rizk.
\newblock {Finding Maximum Cliques on the D-Wave Quantum Annealer}.
\newblock {\em Journal of Signal Processing Systems}, 91(3-4):363--377, 2019.

\bibitem{D-WaveSystems2000QuantumToday}
{D-Wave}.
\newblock {Quantum Computing for the Real World Today}, 2017.

\bibitem{minorminer}
D-Wave.
\newblock {D-Wave Ocean Software Documentation: Minorminer}, 2020.

\bibitem{TechnicalDescriptionDwave}
D-Wave.
\newblock {Technical Description of the D-Wave Quantum Processing Unit}, 2020.

\bibitem{Djidjev2016EfficientAnnealing}
Hristo Djidjev, Guillaume Chapuis, Georg Hahn, and Guillaume Rizk.
\newblock {Efficient Combinatorial Optimization Using Quantum Annealing}.
\newblock {\em LA-UR-16-27928 and arXiv:1801.08653}, 2016.

\bibitem{ErdosRenyi1960}
P.~Erd{\H o}s and A.~R{\'e}nyi.
\newblock {On the Evolution of Random Graphs}.
\newblock {\em Publication of the Math Inst of the Hungarian Academy of
  Sciences}, 5:17--61, 1960.

\bibitem{Hahn2017ReducingBQ}
Georg Hahn and Hristo~N. Djidjev.
\newblock {Reducing Binary Quadratic Forms for More Scalable Quantum
  Annealing}.
\newblock In {\em IEEE International Conference on Rebooting Computing (ICRC)},
  pages 1--8, 2017.

\bibitem{Harris2018}
R.~Harris, Y.~Sato, A.J. Berkley, M.~Reis, F.~Altomare, M.H. Amin, K.~Boothby,
  P.~Bunyk, C.~Deng, C.~Enderud, S.~Huang, E.~Hoskinson, M.W. Johnson,
  E.~Ladizinsky, N.~Ladizinsky, T.~Lanting, R.~Li, T.~Medina, R.~Molavi,
  R.~Neufeld, T.~Oh, I.~Pavlov, I.~Perminov, G.~Poulin-Lamarre, C.~Rich,
  A.~Smirnov, L.~Swenson, N.~Tsai, M.~Volkmann, J.~Whittaker, and J.~Yao.
\newblock Phase transitions in a programmable quantum spin glass simulator.
\newblock {\em Science}, 361(6398):162--165, 2018.

\bibitem{JohnsonNature2011}
Mark Johnson, Mohammad Amin, S~Gildert, Trevor Lanting, F~Hamze, N~Dickson,
  R~Harris, Andrew Berkley, J~Johansson, Paul Bunyk, E~Chapple, C~Enderud,
  Jeremy Hilton, Kamran Karimi, E~Ladizinsky, Nicolas Ladizinsky, T~Oh,
  I~Perminov, C~Rich, and Geordie Rose.
\newblock Quantum annealing with manufactured spins.
\newblock {\em Nature}, 473:194--8, 05 2011.

\bibitem{King2018}
A.~D. King, J.~Carrasquilla, I.~Ozfidan, J.~Raymond, E.~Andriyash, A.~Berkley,
  M.~Reis, T.~M. Lanting, R.~Harris, G.~Poulin-Lamarre, A.~Y. Smirnov, C.~Rich,
  F.~Altomare, P.~Bunyk, J.~Whittaker, L.~Swenson, E.~Hoskinson, Y.~Sato,
  M.~Volkmann, E.~Ladizinsky, M.~Johnson, J.~Hilton, and M.~H. Amin.
\newblock Observation of topological phenomena in a programmable lattice of
  1,800 qubits.
\newblock {\em Nature}, 560:456--460, 2018.

\bibitem{Lucas2014}
A.~Lucas.
\newblock Ising formulations of many {NP} problems.
\newblock {\em Front Phys}, 2(5):1--27, 2014.

\bibitem{McGeoch2018}
Catherine McGeoch.
\newblock {Performance Tuning for D-Wave Quantum Processors}, 2018.

\bibitem{Mockus1974}
Jonas Mockus.
\newblock {On Bayesian Methods for Seeking the Extremum}.
\newblock {\em Optimization Techniques}, pages 400--404, 1974.

\bibitem{Mockus1977}
Jonas Mockus.
\newblock {On Bayesian Methods for Seeking the Extremum and their Application}.
\newblock In {\em IFIP Congress}, pages 195--200, 1977.

\bibitem{Mockus1989}
Jonas Mockus.
\newblock {\em {Bayesian Approach to Global Optimization}}.
\newblock Kluwer Academic Publishers, Dordrecht, 1989.

\bibitem{Mockus2012}
Jonas Mockus.
\newblock {\em Bayesian approach to global optimization: theory and
  applications}.
\newblock Kluwer Academic, 2012.

\bibitem{bayesianopt}
Fernando Nogueira.
\newblock {Bayesian Optimization: Open source constrained global optimization
  tool for Python}, 2020.

\bibitem{Ohkuwa2018}
Masaki Ohkuwa, Hidetoshi Nishimori, and Daniel~A. Lidar.
\newblock Reverse annealing for the fully connected p-spin model.
\newblock {\em Phys. Rev. A}, 98(2):022314, 2018.

\bibitem{Passarelli2020}
Gianluca Passarelli, Ka-Wa Yip, Daniel~A. Lidar, Hidetoshi Nishimori, and
  Procolo Lucignano.
\newblock Reverse quantum annealing of the p-spin model with relaxation.
\newblock {\em Phys. Rev. A}, 101(2):022331, 2020.

\bibitem{optimizing-spin-reversal}
Elijah Pelofske, Georg Hahn, and Hristo Djidjev.
\newblock Optimizing the spin reversal transform on the {D-Wave 2000Q}.
\newblock In {\em 2019 IEEE International Conference on Rebooting Computing
  (ICRC)}, pages 1--8. IEEE, 2019.

\bibitem{Perdomo2011}
A.~Perdomo-Ortiz, S.~E. Venegas-Andraca, and A.~Aspuru-Guzik.
\newblock A study of heuristic guesses for adiabatic quantum computation.
\newblock {\em Quantum Information Processing}, 10:33--52, 2011.

\bibitem{Perdomo2011study}
A.~Perdomo-Ortiz, S.E. Venegas-Andraca, and A.~Aspuru-Guzik.
\newblock A study of heuristic guesses for adiabatic quantum computation.
\newblock {\em Quantum Information Processing}, 10(1):33--52, 2011.

\bibitem{dwave-technology}
D-Wave Systems.
\newblock {Practical Quantum Computing -- D-Wave Technology Overview}, 2020.

\bibitem{Venturelli2019}
Davide Venturelli and Alexei Kondratyev.
\newblock Reverse quantum annealing approach to portfolio optimization
  problems.
\newblock {\em Quantum Mach. Intell.}, 1:17--30, 2019.

\bibitem{Yamashiro2019}
Yu~Yamashiro, Masaki Ohkuwa, Hidetoshi Nishimori, and Daniel~A. Lidar.
\newblock Dynamics of reverse annealing for the fully connected p-spin model.
\newblock {\em Phys. Rev. A}, 100(5):052321, 2019.
\end{thebibliography}

\end{document}